\documentclass[letterpaper]{article} 
\usepackage{aaai2026}  
\makeatletter
\renewcommand{\copyright@text}{
   arXiv preprint
}
\makeatother
\usepackage{times}  
\usepackage{helvet}  
\usepackage{courier}  
\usepackage[hyphens]{url}  
\usepackage{graphicx} 
\DeclareGraphicsExtensions{.pdf,.png,.jpg}
\usepackage{natbib}  
\usepackage{caption} 
\usepackage{algorithm}
\usepackage{algorithmic}
\usepackage{amsmath}

\usepackage{newfloat}
\usepackage{listings}
\usepackage{amssymb} 
\usepackage{epstopdf-base}
\DeclareCaptionStyle{ruled}{labelfont=normalfont,labelsep=colon,strut=off} 
\floatstyle{ruled}
\newfloat{listing}{tb}{lst}{}
\floatname{listing}{Listing}
\title{Large Scale Retrieval for the LinkedIn Feed using Causal Language Models}
\author{
    Sudarshan Srinivasa Ramanujam\equalcontrib,
    Antonio Alonso\equalcontrib,
    Saurabh Kataria\equalcontrib,
    Siddharth Dangi\equalcontrib,
    Akhilesh Gupta\equalcontrib,
    Birjodh Singh Tiwana\equalcontrib,
    Manas Somaiya\equalcontrib,
    Luke Simon\equalcontrib,
    David Byrne\equalcontrib, 
    Sojeong Ha,
    Sen Zhou,
    Andrei Akterskii,
    Zhanglong Liu,
    Samira Sriram,
    Crescent Xiong,
    Zhoutao Pei,
    Angela Shao,
    Alex Li,
    Annie Xiao,
    Caitlin Kolb,
    Thomas Kistler,
    Zach Moore,
    Hamed Firooz\textsuperscript{\rm 1}\equalcontrib
}
\affiliations{
    LinkedIn Corporation, Sunnyvale, CA, USA\\
    \{ sramanujam, aalonso, skataria, sdangi, akgupta ,btiwana, msomaiya, davbyrne, soha, lsimon, sezhou, aakterskii, johliu, sasriram, crxiong, zpei, ashao, alexli, anxiao, ckolb, tkistler, zmoore \}@linkedin.com
    \textsuperscript{\rm 1}Work done while at LinkedIn
}

\usepackage{bibentry}

\begin{document}

\maketitle

\begin{abstract}
In large-scale recommendation systems like the LinkedIn Feed, the retrieval stage is critical for narrowing hundreds of millions of potential candidates to a manageable subset for ranking. LinkedIn's Feed serves suggested content from outside of the member's network (based on the member's topical interests), where 2000 candidates are retrieved from a pool of hundreds of millions candidate with a latency budget of a few milliseconds and inbound QPS of several thousand per second. This paper presents a novel retrieval approach that fine-tunes a large causal language model (Meta’s LLaMA 3) as a dual encoder to generate high quality embeddings for both users (members) and content (items), using only textual input. We describe the end-to-end pipeline, including prompt design for embedding generation, techniques for fine-tuning at LinkedIn's scale, and infrastructure for low latency, cost effective online serving. We share our findings on how quantizing numerical features in the prompt enables the information to get properly encoded in the embedding, facilitating greater alignment between the retrieval and ranking layer. The system was evaluated using offline metrics and an online A/B test, which showed substantial improvements in member engagement. We observed significant gains among newer members, who often lack strong network connections, indicating that high-quality suggested content aids retention. This work demonstrates how generative language models can be effectively adapted for real time, high throughput retrieval in industrial applications.
\end{abstract}


\section{Introduction}
\begin{figure}
    \centering
    \includegraphics[width=0.8\linewidth]{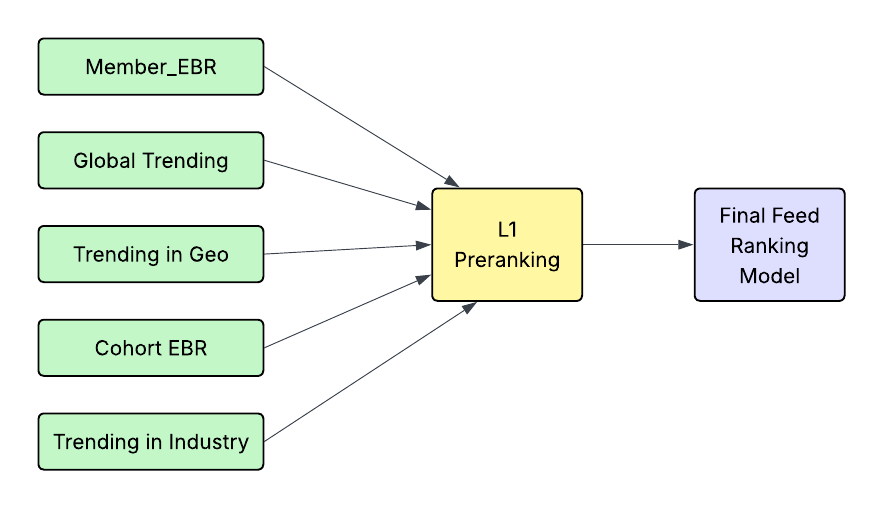}
    \caption{Architecture of the current Suggested Content Ranking and Retrieval Set up. Five sources are shown as examples to illustrate retrieval from multiple sources. In practice, the system has around many more sources} 
    \label{fig:topicality_sources}
\end{figure}

Modern content recommendation systems rely heavily on fast, scalable retrieval techniques to surface relevant items from large candidate pools. At LinkedIn, the Feed retrieval stack has evolved into a highly complex ecosystem comprising multiple index types, including inverted indices of chronologically ordered member activities~\cite{GhikeFollowfeed}, trending sources, collaborative filtering and two-tower embedding-based retrieval (EBR) systems ~\cite{BorisyukCikm2024} for surfacing content from unconnected members. While this multi-index architecture has enabled targeted and personalized Feed experiences, it has also introduced significant engineering complexity and operational overhead, particularly when integrating heterogeneous retrieval signals at scale.
Recent advancements in large language models (LLMs) present a compelling opportunity to re-imagine retrieval architectures~\cite{ZhaiSigir2024}. LLMs pretrained on massive corpora have demonstrated strong capabilities in representation learning and semantic understanding, making them increasingly attractive for retrieval tasks. However, these models are typically not optimized for platform-specific engagement objectives or constrained retrieval latency requirements.
In this work, we propose a novel LLM-based approach to retrieval for the LinkedIn Feed. Our method builds upon an off-the-shelf pretrained LLM, which we further fine-tune using large-scale engagement data specific to LinkedIn. The goal is to directly optimize the model to generate semantically rich query and item representations that capture the nuanced preferences of our member base. We then use these representations in a dense retrieval setup to replace our existing retrieval pipelines (see Figure \ref{fig:topicality_sources} for the current list of sources in the suggested content model).
A key benefit of this approach is the consolidation of disparate retrieval pathways into a unified, embedding-based system. By leveraging a single EBR framework fine-tuned with engagement-supervised LLMs, we can simplify system architecture, reduce maintenance overhead, and enable more coherent ranking stages downstream. Empirical results demonstrate that our approach not only improves relevance compared to multiple retrieval sources with different index structures, but also improves upon latency and throughput.

\subsection{Key Contributions}
\begin{itemize}
    \item We present a practical method for fine-tuning a large language model using real-world engagement signals to enhance retrieval performance in a production-scale feed retrieval setting.
    \item We provide extensive offline and online evaluation to show the effectiveness of our approach in improving relevance and retrieval efficiency for the LinkedIn Feed.
    \item We share our learnings on practical techniques to ensure the important features we added to the prompt are effectively encoded into embeddings.
\end{itemize}

\section{Related Work}
Modern content recommendation systems are typically built upon a multi-stage architecture, commonly involving a retrieval stage followed by a ranking stage. Our work contributes to the evolving landscape of retrieval in recommendation systems, particularly at the intersection of traditional multi-index approaches and the emerging capabilities of Large Language Models (LLMs).
\subsubsection{\textbf{Traditional Retrieval Architectures in Recommendation Systems}}
Early recommendation systems often relied on simpler retrieval mechanisms, such as collaborative filtering (CF) based on user-item interaction histories ~\cite{sarwar2001item,koren2009matrix} or content-based filtering using item metadata. As content platforms scaled, the need for faster and more efficient retrieval led to the adoption of inverted indices for keyword-based search and chronological retrieval for feeds, as seen in social media platforms ~\cite{wang2018understanding}. LinkedIn's Feed retrieval stack exemplifies this evolution, employing a combination of inverted indices for chronologically ordered activities ~\cite{GhikeFollowfeed}, trending sources, and collaborative filtering. This heterogeneous approach allows for diverse signals to contribute to candidate generation, addressing various aspects of user interest (e.g., recency, popularity, similarity to past interactions).
More recently, Embedding-Based Retrieval (EBR) systems, particularly two-tower models, have become a cornerstone of large-scale retrieval ~\cite{covington2016deep, ying2018graph}. These models learn dense vector representations (embeddings) for both queries (users or contexts) and items, enabling efficient nearest-neighbor search in a high-dimensional space. The "two-tower" architecture separates the embedding computation for queries and items, allowing for pre-computation of item embeddings and real-time query embedding, facilitating fast approximate nearest neighbor (ANN) search using techniques like FAISS ~\cite{johnson2019billion} or ScaNN ~\cite{andoni2020scann}. LinkedIn's adoption of two-tower EBR systems~\cite{BorisyukCikm2024} for surfacing content from unconnected members aligns with this industry trend.
\subsubsection{\textbf{Large Language Models in Recommendation and Retrieval}}
The advent of Large Language Models (LLMs) has marked a significant paradigm shift across various NLP tasks, and their application to recommendation systems is a rapidly growing area of research~\cite{covington2016deep,huang2023survey}. LLMs, pre-trained on vast text corpora, possess remarkable capabilities in semantic understanding, contextual reasoning, and representation learning. This has led to their exploration in recommendation for tasks such as generative recommendation~\cite{gao2023generative,liu2024taxonomy}, LLM based feature extraction from text associated with user profiles, item descriptions, and interaction history for retrieval and ranking models \cite{yan2024improving}, and integrating LLM for multi-turn dialogues based recommendations~\cite{feng2023leveraging}.
For retrieval, LLMs offer the potential to create semantically richer embeddings that capture nuanced relationships between users and items, moving beyond simple keyword matching or explicit collaborative signals. Models like BERT~\cite{devlin2019bert} and more recent LLMs have been used as powerful encoders for text, enabling highly effective dense retrieval, where the relevance is determined by the similarity of dense vectors~\cite{karpukhin2020dense}.
\subsubsection{\textbf{Fine-tuning LLMs for Recommendation}}
To address the limitations of off-the-shelf LLMs, a growing body of work focuses on fine-tuning these models for specific recommendation tasks and leveraging domain-specific data. This often involves:
\begin{itemize}
    \item \textbf{Supervised Fine-tuning (SFT)} Training the LLM on labeled datasets of user-item interactions, where the goal is to optimize for explicit engagement signals (e.g., positive vs. negative interactions)~\cite{casalegno2025fine}. This aligns the LLM's representations with the actual behaviors and preferences observed on the platform.
    \item \textbf{Continued Pre-training} Further pre-training LLMs on large-scale unlabeled domain-specific data (e.g., a platform's entire content corpus or user-generated text) to enhance their understanding of the particular domain's vocabulary and concepts~\cite{netflix2025transformer}.
    \item \textbf{Reinforcement Learning from Human Feedback (RLHF)} While more commonly applied to text generation, RLHF principles can be adapted to align LLM-based recommenders with human preferences for relevance, diversity, and quality~\cite{ouyang2022training}.
\end{itemize}


\section{Datasets And Prompt Construction for Finetuning}
\label{sec:prompt_construction}
Our overall optimization target is to increase the number of daily unique "Professional Interactors", which is defined as the number of unique members who take one or "profession interaction" (PI) actions (e.g., long dwell, react, comment, repost, etc.) that contribute to the LinkedIn Feed's Knowledge Marketplace.  Our training data comes from historical Feed interaction logs, and each row in our training dataset is a tuple of (target post features, member features, label), where the member and target post features are feature dictionaries and label is a binary label indicating whether the member took a PI on that post or not.

For the target post, we have the following features:
\begin{itemize}
\item type of the post (original post, group post, like/comment on a previous post)
\item what is being shared (text, image, video, job change, etc.)
\item author information (author name, profile headline, company, industry, title)
\item post popularity features (\# of times the post has been liked, viewed for more than T secs, etc.)
\item article title/source (if the post contains an article link)
\item text of the post
\end{itemize}

The member features include the following: name, profile headline \& summary, industry, skill(s), location, job and education history, certifications, and languages spoken.  For each member, we also have their ``activity history sequence" -- a time-ordered list of Feed posts on which the member has previously taken a PI. Note that we did perform several ablation studies for  what posts to include in this history sequence \cite{firooz2025360brew}:
\begin{itemize}
\item All posts (i.e., both positive and negative engagements)
\item Only positive PI engagements
\item No history (i.e., only include member profile information)
\end{itemize}
These ablation study results will be presented in the Results section. In the end, including only positive PI engagement in the activity history sequence performed the best, and is the method that we used.

A ``prompt library" is used to convert these features into a target post prompt and member prompt, respectively.

The text of the target post prompt looks like:
\begin{center}
\begin{minipage}{0.8\linewidth}
\begin{verbatim}
<ST_P1>Post feature 1
<ST_P2>Post feature 2
...
<ST_PN>Post feature N

\end{verbatim}
\end{minipage}
\end{center}
The text of the member prompt looks like:
\begin{center}
\begin{minipage}{0.8\linewidth}
\begin{verbatim}
<ST_M0>System prompt
<ST_M1>Member feature 1
<ST_M2>Member feature 2
...
<ST_MN>Member feature N
<ST_history>
  <ST_history_post><post 1 text>
  <ST_history_post><post 2 text>
  ...
  <ST_history_post><post H text>
  
\end{verbatim}
\end{minipage}
\end{center}

The \verb|<post 1 text>, ..., <post H text>| inside of the member prompt are generated similarly to the target post's prompt. The number of history posts that fit into the prompt is dynamic based on the total maximum context length (20,480 in our experiments) and the tokenized length of each post.

The system prompt is the following: ``You are provided with a member's profile information, along with a set of historical feed posts that the member engaged with. Your task is to analyze the historical engagement data along with the member profile."

Note that the \verb|<ST_...>| strings above are special token strings that are added to the tokenizer's vocabulary so that they get tokenized as a single token, in order to reduce the size of the prompts. The strings shown above are just examples for explanatory purposes -- the actual strings are not shared here for security purposes to discourage any prompt injection attacks.

As shown in Figure \ref{fig:model_arch_overview}, the target post prompt and the member prompt are fed as input to an LLM in order to generate the target post and member embedding, respectively.

\section{Modeling Architecture}
\label{sec:modeling_arch}
\begin{figure}
    \centering
    \includegraphics[width=0.8\linewidth]{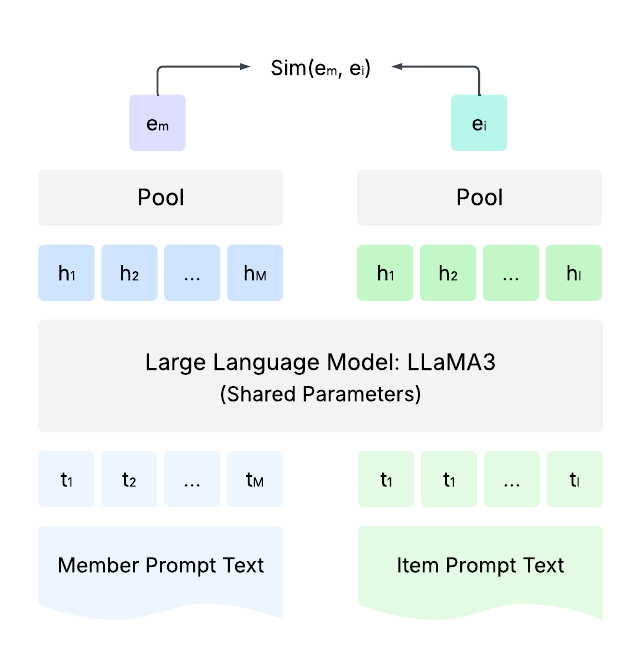}
    \caption{Dual encoder architecture utilizing a shared LLM for text-based retrieval. Member and item texts are processed separately through the LLM, generating token-level hidden representations. A pooling layer aggregates these representations into embeddings, which are then compared using a similarity function.} 
    \label{fig:model_arch_overview}
\end{figure}

We leverage a pre-trained, decoder-only transformer-based causal language model as our initial base model (Meta LLaMA-3 \cite{dubey2024llama}). We optimize the base LLM for embedding-based retrieval (EBR) through fine-tuning using a dual-encoder architecture (with a single shared LLM) that is used to encode both members and items into a shared embedding space. Figure \ref{fig:model_arch_overview} depicts a high level illustration of the architecture.

\subsubsection{\textbf{Embedding Generation}}
\label{subsec:embgen}
Both member and item texts undergo tokenization before being processed by the LLM. Given a tokenized sequence $t$ of length $L$, the LLM produces a sequence of hidden states, $H \in \mathbb{R}^{L \times d}$ where $d$ denotes the dimensionality of the hidden states. A pooling function is subsequently applied to generate a fixed-dimensional, dense representation. Specifically, the embedding for a member token sequence $t_m$ is computed as: $e_m = \textit{pool}(H_m)$. An analogous process is used to generate an item embedding $e_i$. We discuss pooling functions further in subsection \ref{subsec:pooling}.

\subsubsection{\textbf{Measuring Member-Item Similarity}}
The similarity between member and item embeddings is quantified using a similarity function, $s(e_m, e_i)$. In this study, cosine similarity is employed: $s(e_m, e_i) = \frac{e_m \cdot e_i}{||e_m|| \cdot ||e_i||}$. This similarity score serves as the primary retrieval ranking metric, enabling efficient identification of the most relevant items for a given member.

\subsection{Pooling}
\label{subsec:pooling}
The pooling function aggregates the token-level hidden states into the fixed-dimensional, dense embeddings. 

\textbf{Mean Pooling:} Given an input sequence consisting of \(L\) tokens with hidden states \(H \in \mathbb{R}^{L \times d}\) (with \(H_i\) as the \(i\)-th token), the pooled embedding is
\[
e = \frac{1}{L} \sum_{i=1}^{L} H_i.
\]
This method yields a holistic representation by averaging over tokens.
In this work, we considered mean pooling various sequence lengths to land on the best performing model and results are discussed in subsection \ref{subsec:last_n_pooling}

\subsection{Training objectives}
\label{subsec:losses}
When fine-tuning for the retrieval task, we wish to optimize an objective where embeddings of positive member-item pairs are drawn closer together, while negative  pairs are pushed apart in the embedding space. For training, we leverage binary data that captures whether a positive or negative action was taken between a member and an item. In this setting, each member--item pair is assigned a binary label \( y \in \{0,1\} \). We explore the following loss functions that leverage the labelled data to effectively learn embedding similarities.

\textbf{Binary Cross-Entropy (BCE)}:
In this formulation, the similarity score \( s(e_m, e_i) \) is scaled by the temperature \(\tau\) and interpreted as a logit. The corresponding probability is computed using the sigmoid function \(\sigma(\cdot)\):
\[
P(y=1 \mid e_m, e_i) = \sigma\left(\frac{s(e_m, e_i)}{\tau}\right).
\]
The BCE loss is then expressed as:

\begin{equation}
\begin{split}
L_{\text{BCE}} = - \Big[ y \log \sigma\left(\frac{s(e_m, e_i)}{\tau}\right) \\
+ (1 - y) \log \left(1 - \sigma\left(\frac{s(e_m, e_i)}{\tau}\right)\right) \Big]
\end{split}
\end{equation}

\textbf{InfoNCE}:
For a given member embedding \( e_m \) and a corresponding positive item embedding \( e_i^+ \), along with a set of negative item embeddings \( \{e_i^-\} \), the InfoNCE loss \cite{oord2018representation}  is defined as:

\begin{equation}
L_{\text{InfoNCE}} = 
- \log 
\frac{
    \exp\!\left(\frac{\text{s}(e_m, e_i^+)}{\tau}\right)
}{
    \exp\!\left(\frac{\text{s}(e_m, e_i^+)}{\tau}\right)
    + 
    \sum_j \exp\!\left(\frac{\text{s}(e_m, e_{i_j}^-)}{\tau}\right)
}
\label{eqn:infonce}
\end{equation}
where \(\text{s}(\cdot,\cdot)\) denotes the similarity function used, and \(\tau\) is a temperature parameter. This loss encourages the similarity of positive pairs to be higher than that of negative pairs by emphasizing relative ranking. InfoNce is more commonly used for retrieval and we expected this to perform better. Binary Cross Entropy Loss served as a good baseline for us to compare the performance of the model with the infoNce loss.




\subsection{Easy and Hard Negative Mining}

The negatives used for the InfoNCE loss described in Equation \ref{eqn:infonce} are a combination of easy and hard negatives. Following Google’s two-tower retrieval literature~\cite{DBLP:conf/www/YangYCHLWXC20}, we mix easy and hard negatives. Mixed Negative Sampling (MNS) combines batch (in-batch) negatives with uniformly sampled corpus negatives to reduce selection bias in implicit feedback data and has shown offline and online gains in large-scale production (e.g., Google Play). Our setup extends this idea by adding per-member hard-negative mining on top of in-batch sampling.

In each training step, in-batch negatives are sampled from the global mini-batch (as opposed to the local mini-batch on each individual GPU). These examples provide weak negative pairs, artificially creating impressions with no action, improving training stability and increasing the number of training examples seen by a factor of batch size$^2$. \newline
We also built tunable parameters to dynamically sample hard negatives per batch for each individual member, in addition to the aforementioned easy negatives. Operationally, our sampler mirrors triplet training: for each anchor–positive, we mine $K$ hard negatives (near-miss impressions for that member) and sample 
$J$ easy negatives (global in-batch). Triplet loss maximizes a margin between the anchor–positive and anchor–negative; InfoNCE replaces the margin with a softmax over the pooled negatives, which often yields smoother optimization at scale~\cite{DBLP:conf/cvpr/SchroffKP15}.

\begin{itemize}
    \item \textbf{Easy Negatives} Negatives sampled from  the global batch (across all GPUs)
    \item \textbf{Hard Negatives} Items which were impressed by a member without an engagement action. We create this map offline and store it in memory for sampling required number of hard negatives at training time. 
\end{itemize}

\subsection{Matryoshka Embeddings}
\label{subsec: MRL}
As part of our training procedure, we also employ Matryoshka Representation Learning (MRL) \cite{Kusupati2022MatryoshkaRL}, a framework that learns nested, size-adaptive representations by optimizing multiple sub-representations simultaneously. At each level, progressively larger subsets of the embedding are encouraged to capture increasingly rich and informative features. This property is advantageous in production environments, where models must often operate under varying computational or memory constraints. By ensuring that smaller sub-representations remain effective for the downstream task, MRL enables flexible, efficient deployment without the need for retraining or architecture modifications. The learning process is guided by the Matryoshka Loss, defined as:

\begin{equation}
\mathcal{L}_{\text{MRL}} = \sum_{k=1}^{K} \lambda_k \cdot \mathcal{L}_k
\end{equation}

where \( K \) denotes the total number of representation sizes, \( \lambda_k \) are weighting coefficients, and \( \mathcal{L}_k \) is the task-specific loss computed using only the first \( k \) dimensions of the representation. In our use case, we average the infoNce loss that is computed for each of first 'k' dimensions.

\section{Offline Evaluation}
We evaluate our retrieval model using \textbf{Recall@$k$}, treating the final ranking model as the oracle. The evaluation proceeds as follows:  

\begin{enumerate}
    \item Randomly sample a set of unique members who have had engagement in the LinkedIn Feed.
    \item For each member, rank all $N$ items from their aggregated sessions using the final ranking model, and save the top $n$ suggested content items in $Set_1$.
    \item Rank the top $n$ items per member using similarity scores from the retrieval model (e.g., cosine similarity between query and item embeddings), and save the top $k$ items in $Set_2$.
    \item Compute Recall@$k$ as:  
    \[
        \mathrm{Recall}@k = \frac{|Set_1 \cap Set_2|}{|Set_1|}
    \]  
    where $Set_1$ contains the top-$n$ items according to the ranking model, and $Set_2$ contains the top-$k$ items according to the retrieval model.
    \item Average the Recall@$k$ values across all members to obtain the final metric.
\end{enumerate}

We computed recall at the member level rather than the session level because the number of suggested content items viewed within a single session is typically very small. This limited interaction scope may not provide a comprehensive assessment of the performance of the retrieval model. By evaluating recall over all suggested content items that a member has engaged with within a defined time period (e.g., one week), we obtain a more holistic measure of the model’s ability to retrieve relevant content across multiple interactions. This approach aligns with our objective of developing a retrieval model that effectively serves the interests of our members.

\section{Online System}
For running online retrieval for serving member queries, we built multiple workflows, as we detail next.

\begin{figure}
    \centering
    \includegraphics[width=0.8\linewidth]{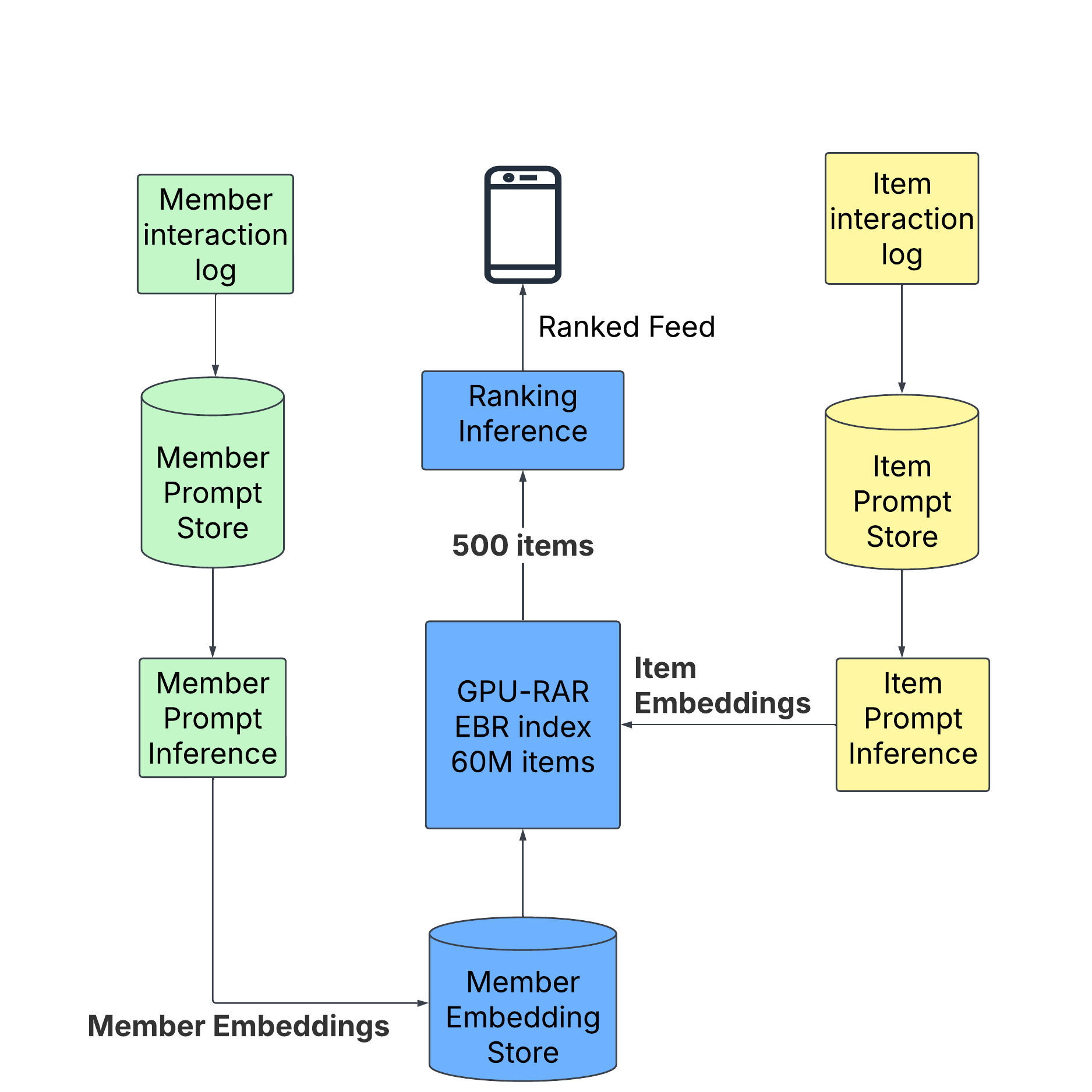}
    \caption{Online System for the Retrieval and Ranking Model}
    \label{fig:online_deployment}
\end{figure}

\subsubsection{\textbf{Nearline Item \& Member activity log generation}}
When items or member profiles are created or updated on LinkedIn, we capture those triggers into an item and member activity log. When members interact with items (e.g. like, comment, share, etc.), we capture those interactions too by processing tracking data from clients into the item and member activity logs. We minimize the latency in capturing these interactions by using direct RPC calls for service-to-service communication, where possible, over nearline stream processing.
\subsubsection{\textbf{Nearline Item \& Member prompt generation}}
Next, we process the item and member activity logs into corresponding item and member prompts using pre-defined prompt templates where data such as item text, member profile information, and item popularity counts are fetched and populated into the templates to construct prompts that include interaction history. This ensures freshness of prompt data. Finally, we push these fully decorated prompts to a key-value store for online access during retrieval and to a nearline stream processor for generating embeddings.
\subsubsection{\textbf{Nearline Item \& Member embedding generation using online LLM inference}}
\label{subsec:emb_gen}
We feed updated item prompts for each item creation and item update into a LLM inference server hosting our fine-tuned LLM to generate embeddings as described in Section \ref{subsec:embgen}. We ingest the generated item embeddings into a GPU index for online kNN (k nearest neighbor) retrieval \cite{BorisyukCikm2024}. The GPU index currently lets us define a custom pytorch model in order to do the kNN operation. For this specific use case, we use a simple cosine similarity model and the top 'k' documents with the highest cosine similarity scores with the member embedding are retrieved from the index. We refer to this as a "GPU Retrieval as Ranking" (GPU-RAR) index.
Similarly, we feed updated member prompts for each member profile creation and member interaction activity to the LLM inference server to generate embeddings and ingest them into an online key-value store for access during online retrieval.
This embedding generation process is done using nearline stream processing to control the LLM inference rate by batching updates in configurable window sizes because our scale results in thousands of input prompt updates per second. We tradeoff GPU compute used against embedding freshness by using shorter window sizes for increased freshness which helps capture evolving member interests and item popularity.  We ensure that newly created items are indexed in the GPU-RAR index within a minute of creation and interactions on existing items result in an update of their embeddings within 30 minutes. Similarly, newly added member profiles are captured in their member query embeddings within a minute and activities by existing members result in an update of their embeddings within 30 minutes.
\subsubsection{\textbf{Online GPU-RAR kNN retrieval with attribute-based matching}}
To serve an online LinkedIn feed query for a member, we fetch member query embeddings and run online kNN against the GPU-RAR item embeddings index to retrieve top K items, while applying business logic filtering and privacy rules, that are then sent to the ranker layer. We apply filters as part of the kNN query to ensure that the selected items are approved by our trust classifiers, match the languages understood by the viewer, not authored by members blocked by the viewer, and not already seen by the viewer. The pre-computation of embeddings allows us to achieve sub-50ms retrieval latency for serving tens of thousands of queries per second on a corpus of hundreds of millions of items while maintaining embedding freshness of a few minutes.

\section{Implementation Details}
We used 5M member-item pairs from public engagement in the LinkedIn Feed as our training samples.  We used 8 H100 GPUs for each training run with a per-GPU batch size of 4. We experimented with both Meta-LLaMA 3B and 1B parameter models with various combinations of Matryoshka Embeddings.

In the online stack, we used a cluster of 48 H100 GPUs for nearline item and member embedding inference. This cluster also handled the embedding inference traffic generated for back-filling the embeddings for new experimental models for the entire item and member corpus. We used a cluster of 24 GPUs for indexing item embeddings and for performing online GPU-RAR kNN retrieval with attribute-based matching. We used a retrieval model that employed cosine similarity between member and item candidates to get the top 1000 candidates to feed to subsequent layers of the ranking stack. 

\section{Results}
The Meta LLaMA-3 model with 3B parameters was leveraged as the base model for further fine-tuning to get all the results discussed in this section. This model by default has an output dimension of 3072 unless specified otherwise.

\subsection{Dual Encoder Results}
\begin{table}[ht]
\centering
\small
\begin{tabular}{|p{4.2cm}|p{1.5cm}|c|}
\hline
\textbf{Loss}                          & \textbf{Recall@10} \\ \hline
Random                                 & 0.0700             \\ \hline
Llama-3 Without Finetuning             & 0.2434             \\ \hline
BCE                                    & 0.3944             \\ \hline
InfoNCE                                & 0.4238             \\ \hline
InfoNCE with Matryoshka Loss (full 3072 dim)           & 0.4242             \\ \hline
\end{tabular}
\caption{Results for the Dual Encoder architecture under different training objectives and using all dimensions}
\label{dual_encoder_result_table}
\end{table}

From Table \ref{dual_encoder_result_table}, we observe that the InfoNCE loss is able to outperform BCE loss. We also observe that using Matryoshka Representation Learning to learn multiple dimensionalities together does not hurt overall performance.



\subsection{Dual Encoder Results with Matryoshka Learning}

One of the key motivations to do Matryoshka learning was to reduce the dimensionality of the final embedding. This enables us to reduce the storage cost in the GPU index from which documents are retrieved in the online system. We trained a model with the method described in subsection \ref{subsec: MRL} and evaluated multiple embedding dimensions in parallel.
\begin{table}[ht]
\centering
\small
\begin{tabular}{|p{3.5cm}|p{1.5cm}|c|}
\hline
\textbf{Embedding Dimensions}   & \textbf{Recall@10} \\ \hline
3072 (all dims)                 & 0.4242             \\ \hline
2048                            & 0.4248             \\ \hline
1024                            & 0.4237             \\ \hline
512                             & 0.4225             \\ \hline
50                              & 0.3716             \\ \hline
\end{tabular}
\caption{Evaluating performance on reducing embedding size after training with Matryoshka Loss.}
\label{recall_vs_dim_table}
\end{table}
\begin{figure}[!h]
    \centering
    \includegraphics[width=0.85\linewidth]{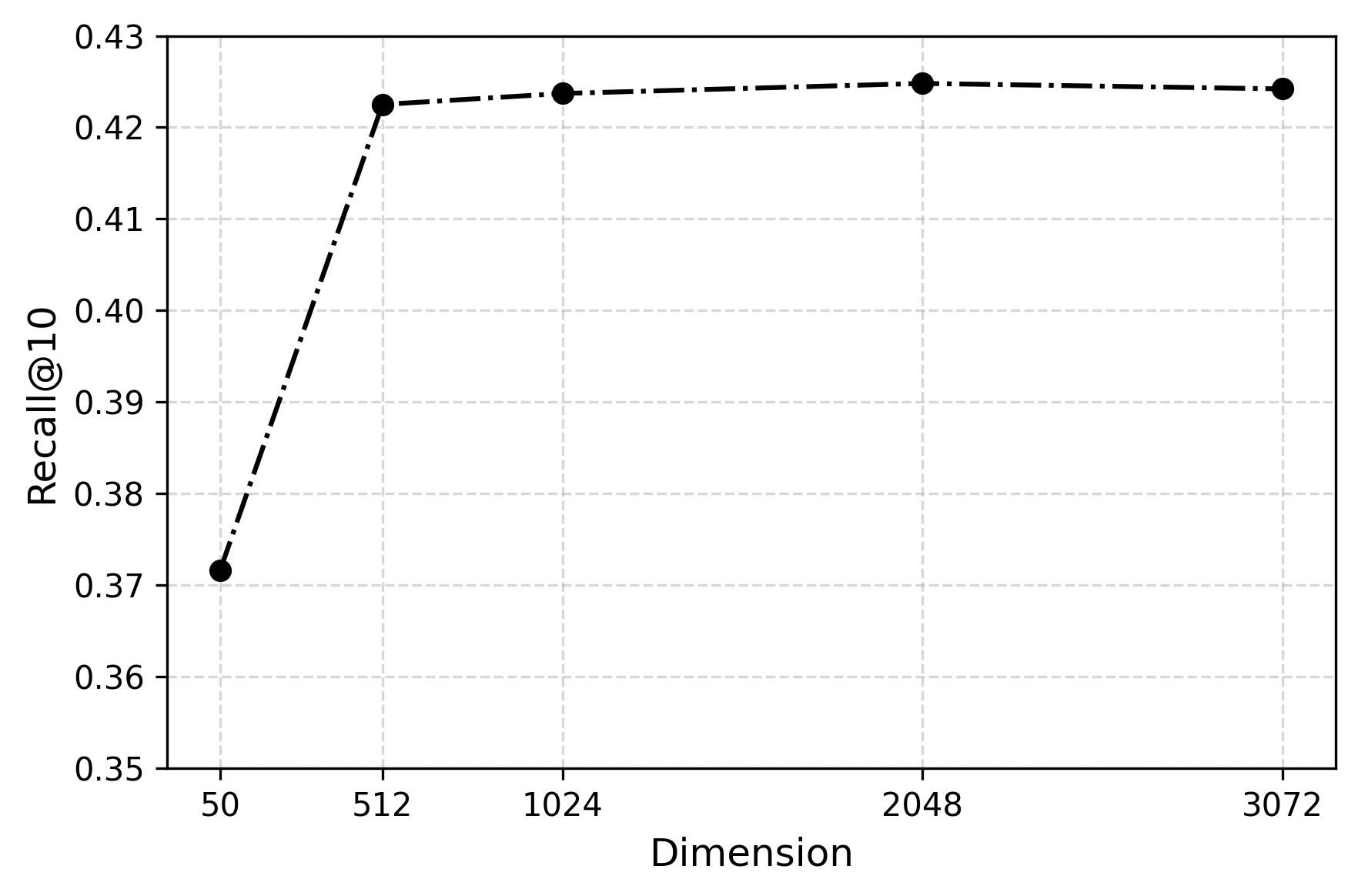}
    \caption{Results for Recall@10 versus Embedding Dimension when training using MRL.}
    \label{fig:recall_vs_dim}
\end{figure}

The plot in Figure \ref{fig:recall_vs_dim} demonstrates that lowering the dimension to 512 does not significantly impact the recall numbers and offers a lot of potential to reduce the storage size of the final embeddings.

\subsubsection{Comparison of Matryoshka Learning against directly lowering dimension using MLPs}

As illustrated in Table \ref{table: dim_red_vs_mrl}, it is possible to fine-tune a model with lower dimensional embeddings by adding MLPs to the final layer. We experimented with 2 methods: 1) pooling and then using an MLP to reduce the dimension, and 2) reducing the dimension before pooling. 
From the results(recall@10), we observed that MRL outperformed pooling to a specific dimension using MLPs.

\begin{table}[ht]
\small
\begin{tabular}{|p{1.5cm}|p{1.75cm}|p{1.75cm}|p{1.5cm}|}
\hline
\textbf{Dimension} & \textbf{Pre-pool} & \textbf{Post-pool} & \textbf{MRL} \\ \hline
2048      & -3.2\%              & -2.7\%               & \textbf{+0.1\%} \\ \hline
1024      & -2.7\%              & -2.3\%               & \textbf{-0.2\%} \\ \hline
512       & -1.1\%              & -1.0\%               & \textbf{-0.8\%} \\ \hline
\end{tabular}
\caption{Percentage difference in Recall@10 for different embedding dimension sizes compared to using the full 3072 dimensions. We consider a pre-pooling projection (project to desired dimension then pool embeddings), post-pooling projection (pool embeddings and then project to desired dimension) and applying MRL.}
\label{table: dim_red_vs_mrl}
\end{table}


\subsection{Effect of Hard Negatives on Recall Metrics}
\begin{table}[ht]
\centering
\small
\renewcommand{\arraystretch}{1.1} 
\begin{tabular}{|l|c|}
\hline
\textbf{Hard Negatives} & \textbf{Recall@10} \\ \hline
Easy negatives only & Baseline \\ \hline
Easy negatives + 1 hard negative/member & +2.0\% \\ \hline
Easy negatives + 2 hard negatives/member & +3.6\% \\ \hline
\end{tabular}
\caption{Effect of adding per-member hard negatives on Recall@10.}
\label{hard_negatives_result_table}
\end{table}

Our experiment results clearly demonstrate that adding hard negatives helps to improve recall metrics. We are exploring ways to introduce more hard negatives without hurting recall metrics too much. Some ideas currently being worked on are progressively increasing number of hard negatives during training, curriculum learning by bumping hard negatives in an iterative way

\subsection{Effect of Filtering for positives in interaction history}
\begin{table}[htbp]
\centering
\small
\begin{tabular}{|p{1.2cm}|p{3.2cm}|p{3.2cm}|}
\hline
\textbf{Loss} & \textbf{Full History Recall@10} & \textbf{Pos-Only History Recall@10} \\ \hline
BCE           & 0.307                           & 0.3944                              \\ \hline
InfoNCE       & 0.398                           & 0.4238                              \\ \hline
\end{tabular}
\caption{Recall@10 using different member history strategies.}
\label{history_filtering_table}
\end{table}

When we removed all negative engagements from the prompt history and instead extended more positive items that were engaged on, we saw a significant improvement in recall metrics (Table \ref{history_filtering_table}). In all subsequent iterations, negative interactions were removed from the member prompt. 

\subsection{Effect of Last n pooling}
\label{subsec:last_n_pooling}

We share our findings on pooling different set of last 'n' tokens and compare it against pooling all tokens in the last layer in this section: 

\begin{table}[h]
\centering
\small
\begin{tabular}{|c|c|}
\hline
\textbf{Last\_n tokens used} & \textbf{Percent} \\
\hline
All   &   Baseline \\
\hline
500   & -7.0603\% \\
\hline
250   & -5.0040\% \\
\hline
50    & -11.2637\% \\
\hline
1     & -12.3594\% \\
\hline
\end{tabular}
\caption{Percent change across different Last\_n values.}
\label{table:last_n_mean_pool}
\end{table}

From Table \ref{table:last_n_mean_pool}, we noticed that the baseline (pooling the embeddings of all the tokens) performed the best. Pooling any other subset resulted in recall@10 drops. 

\subsection{Can LLMs capture count features that are important for Ranking}

In the course of our work, we came up with a framework with which we could enable the final embeddings to encode important count-based features that are used in the final ranking model. This was critical for the overall success of the retriever and we will share some of our learnings in this section. One of the most important metrics for the retriever is how well it aligns with the ranking model. Hence, if some of the important features for the ranking model cannot be captured in the embedding space, this would result in subpar candidates sent to the ranking model resulting in engagement drops. 




\subsubsection{Results of adding engagement rate to input text and truncating feed post length}

One of the key lessons we learned was that quantizing the count features and passing them to the input prompt ensured better correlation between the final cosine similarity score and the feature value. We demonstrate this with a simple study we did using the item popularity counts, which is an important feature for the final ranking model using a smaller dataset.  

\begin{itemize}
    \item For the \textbf{Baseline} in Table \ref{popularity_correlation_table}, inputs had the entire post text and the raw popularity counts of the posts in the engagement history of members (query side) and in the post text (candidate side). Popularity counts could go to pretty large values (tens of thousands). 
    \item For the \textbf{Candidate Model} in Table \ref{popularity_correlation_table}  all post texts in the query and item side were truncated to the first 60 tokens instead of considering the whole post and popularity rates (range of 1-100 percent) were added as features on top of raw counts.
\end{itemize}

\begin{table}[h]
\centering
\small
\begin{tabular}{|p{2.5cm}|p{1.5cm}|c|c|}
\hline
\textbf{Model} & \textbf{Correlation between popularity counts and cosine similarity scores} & \textbf{Recall@10} \\
\hline
Baseline & \centering -0.0037 & 0.158 \\
\hline
Candidate Model & \centering 0.1156 & 0.1839 \\
\hline
\end{tabular}
\caption{Results for measuring correlation between retrieval scores and popularity percentages}
\label{popularity_correlation_table}
\end{table}
The column titled Correlation in Table \ref{popularity_correlation_table} refers to correlation between item popularity counts and cosine similarity score between the embeddings.
From the results in Table \ref{popularity_correlation_table}, there is evidence (improved correlation and recall improvements by 15\%)that adding count-based features important for the ranking model in a quantized manner to the input prompt can be captured by the embeddings we generate for retrieval. This aligns with the fact that the Llama-3 tokenizer groups 3 digits together as a single token \cite{arnett2024llama3_tokenizer}. Since the counts were now captured by a single token, this could be an added factor in the feature correlation getting better.

\subsection{Online Ramp Results}

\subsubsection{}

We used all the learnings from the offline experimentation and used the following set of hyperparameters to prepare a model for the online ramp: 

\begin{itemize}
    \item Llama 3B  base model with 3072 dimensions in last layer
    \item Use only positive interactions in member prompt history
    \item Mean pool embeddings of all tokens
    \item Employed 2 hard negatives for a per GPU batch size of 4
    \item Quantized popularity features to percentages
\end{itemize}

We A/B tested the new LLM-based retriever against the existing suggested content retrieval model (Figure \ref{fig:topicality_sources}) and the final ranking stages were identical for both setups.

We observed significantly improved suggested content recommendations, which were served to our members, and as expected we impacted members with few connections and new members to LinkedIn substantially more in addition to overall metric improvements online.

\begin{itemize}
    \item \textbf {Revenue} Increased by \textbf{+0.8\%} (pval of 0.03) as a result of increased scroll depth and member interaction in the LinkedIn Feed
    \item \textbf{Daily Unique Professional Interactors} Increased by \textbf{+0.2\%} (pval of 0.005) in the LinkedIn feed
\end{itemize}

Inspecting the metrics among member cohorts with lower liquidity of candidate posts and fewer connections to other members, we observed the following metric impacts:

\begin{itemize}
    \item \textbf{Daily Active Unique Users} Increased by \textbf{+0.23\%} (pval of 0.05)
    \item \textbf{Daily Unique Professional Interactions} Increased by \textbf{+1.17\%} (pval less than 0.0001)
    \item \textbf{Revenue} Increased by \textbf{+3.29\%} (pval of 0.03) for this user group, indicating that the majority of platform-wide impact came from infrequent members and members with fewer connections, for whom suggested content plays a much more vital role.
\end{itemize}

\section{Conclusion}
In this paper, we presented the redesign and implementation of a modern retrieval system for the LinkedIn feed. We explored modeling choices, loss functions, and pooling strategies, and demonstrated through offline experiments and online A/B tests that fine-tuning large language models can substantially improve the quality of recommended content for our members.

Looking ahead, we plan to improve handling of unimpressed content at the retrieval stage to encourage exploration, and to investigate more effective distillation strategies for deriving dual encoders from cross encoders. Building on the efficacy of matryoshka learning, we are working on reducing embedding dimensionality to lower storage costs. We are also exploring LLM-powered embeddings for content generated by members’ connections, which represents the bulk of impressed items in the LinkedIn feed.

On the efficiency side, we are investigating methods to shorten input sequences to improve GPU throughput in our nearline system. Finally, we have begun prototyping user-prompt–driven feed recommendation as a re-ranking layer, which is already showing promising early results.

\section*{Acknowledgements}

This work represents the joint efforts across multiple teams in LinkedIn without whom this would not have been possible. We would like to thank (in alphabetical order) Adrian Englhardt, Aman Gupta, Ata Fatahi Baarzi, Bhargav Patel, Christine Lin, Dre Olgiati, Frank Shyu, Ganesh Parameshwaran, Ghulam Ahmed Ansari, Hemeng Tao, Hristo Danchev, Jashua Gupta, Kirill Talanine, Lars Hertel, Mingzhou Zhou, Mohit Kothari, Qingquan Song, Samira Sriram, Samaneh Moghaddam, Steven Shimizu, Sundararaman Ramachandran,  Tejas Dharamsi, Tim Chao, Tim Jurka, Vignesh Kothapalli, Ying Xuan, Youngchae Kim and Yun Dai for supporting this work.

\bibliography{aaai2026}

\end{document}